\documentclass[12pt]{article}
\usepackage{amsmath,amsfonts,url}
 \makeatletter

 \DeclareRobustCommand\em
        {\@nomath\em \ifdim \fontdimen\@ne\font >\z@
                       \upshape \else \slshape \fi}
 \newcommand{\eqlabel}[1]{\label{eq:#1}}
 \newcommand{\Eq}[2][]{\def\t@mp{#1}%
\begin{equation}#2\ifx\t@mp\@empty\notag\else\eqlabel{#1}\fi\end{equation}}
 \newcommand{\Eqaligned}[2][]{\def\t@mp{#1}%
\begin{equation}\begin{aligned}#2\end{aligned}
\ifx\t@mp\@empty\notag\else\eqlabel{#1}\fi
\end{equation}}
 \newcommand{\eq}[1]{(\ref{eq:#1})}	
 \makeatother

\def\H{\mathcal{H}}
\def\M{\mathcal{M}}
\def\Pr{\mathrm{Pr}}
\def\Tr{\mathrm{Tr}}
\def\Iden{\mathbf{I}}
\def\halmos{\quad\vrule height1.1ex width1.1ex depth0ex}

\date{}

 \author{Lev B.\ Levitin and Tommaso Toffoli\\{\small\sl ECE Dept., Boston
University, Boston, MA 02215}}
 \title{Information between quantum systems\\via POVMs}

\begin{document}

\maketitle

\begin{abstract} 

\noindent The concepts of conditional entropy and information between
subsystems of a composite quantum system are generalized to include arbitrary
indirect measurements (POVMs). Some properties of those quantities differ
from those of their classical counterparts; certain equalities and
inequalities of classical information theory may be violated.

\smallskip\noindent
 {\bf Keywords}: Quantum information; quantum conditional entropy; POVM;
quantum relative entropy; entropy defect. PACS 03.67.-a.

\end{abstract}

\noindent In \cite{levitin1,levitin2}, concepts of conditional entropy and
information were introduced for quantum systems with respect to direct (von
Neumann) measurements performed over subsystems of a composite quantum
system.  In this paper the concepts are generalized to include arbitrary
indirect measurements (POVMs). The concepts of `conditional entropy' and
`information' retain their validity for quantum systems, but their properties
differ somewhat from those of their classical counterparts; specifically,
some equalities and inequalities of classical information theory are in
general violated.

\bigskip

Consider a composite quantum system consisting of two subsystems $A$ and $B$,
the Hilbert space $\H$ of the system being the tensor
product, $\H_A\otimes\H_B$, of the Hilbert spaces of its two
subsystems. The state of the system and the states of its subsystems are
described, respectively, by the joint density matrix $\rho(A,B)$ and the
marginal density matrices $\rho(A)$ and $\rho(B)$. The joint entropy of the
system and the marginal entropies of the two subsystems are, respectively,
$H(A,B)$, $H(A)$, and $H(B)$.

\medskip

Now, let $\M_A =\{M_a(A)\}$ be a countable set of self-adjoint nonnegative
definite operators that form a resolution of the identity (in general,
non-orthogonal) in $\H_A$, and $\M_B =\{M_b(B)\}$ a similar set of operators
in $\H_B$.  The sets $\M_A$ and $\M_B$ correspond to indirect measurements
(POVMs) performed respectively over the systems $A$ and $B$. By Naimark's
theorem\cite{naimark}, any POVM is equivalent to a direct (von Neumann)
measurement performed in an extended Hilbert space. Henceforth we will
consider POVMs that correspond to measurements of a complete set of variables
(represented by a complete set of orthogonal one-dimensional projectors) in
the extended Hilbert space.\footnote
 {The expression ``complete set of variables'' is used here in exactly the
same meaning as ``complete set of physical quantities'' in
\cite[p.\,5]{lifshitz}, namely, as a maximum set of simultaneously measurable
quantum variables (observables). ``Complete set of projectors'' means that
they form a resolution of the identity. Also, since sets $\M_A$ and $\M_B$ are
countable, they correspond to measurements of variables with discrete
spectrum.}

Denote by $\alpha$ and $\beta$ the two random variables that are the results
of measurements $\M_A$ and $\M_B$. The probability distributions of $\alpha$
and $\beta$ are
 \Eqaligned[pralphabeta]{
 \Pr\{\alpha=a\}	&=\Tr\{\rho(A)M_a(A)\},\\
 \Pr\{\beta=b\}		&=\Tr\{\rho(B)M_b(B)\},\\
 \Pr\{\alpha=a,\beta=b\}&=\Tr\left\{\rho(A,B)[M_a(A)\otimes M_b(b)]\right\}.
 }
 
\medskip\noindent{\sc Lemma 1.}\quad For any choice of $\M_A$ and $\M_B$,
 \Eqaligned[lemma1]{
 H(A,B)		&\leq H(\alpha,\beta),\\
 H(A)		&\leq H(\alpha),\\
 H(B)		&\leq H(\beta).
 }
 {\bf Proof.}\quad Inequalities \eq{lemma1} follow from Klein's
lemma\cite{klein} by use of Naimark's theorem.\halmos

\medskip

The conditional density matrix of subsystem $A$ given the result of a
measurement performed over subsystem $B$ can be defined for POVMs in a
similar way as it is defined for von Neumann measurements, namely, following
\cite{balian},
 \Eq[balian]{
 \rho(A|\beta=b) = \frac{\Tr_B\left\{\rho(A,B)[\Iden(A)\otimes M_b(B)]\right\}}
		   {\Tr\left\{\rho(A,B)[\Iden(A)\otimes M_b(B)]\right\}},
 }
 where $\Iden(A)$ is the identity operator in $\H_A$. Note that the
denominator in \eq{balian} is just the probability $\Pr\{\beta=b\}$ for $\beta$ to take on
value $b$.

Then the conditional entropy of system $A$, given measurement $\M_B$
performed on $B$, is
 \Eq[TrTr]{
 H(\!A|\beta)=-\!\sum_b\Tr\!\left\{\rho(\!A,B)[\Iden(A)\otimes M_b(B)]\right\}
	\Tr\left\{\rho(\!A|\beta=b)\log\rho(\!A|\beta=b)\right\}.
 }

By Klein's lemma, for any $\alpha$ and $\beta$ (i.e., for any measurements
$\M_A$ and $\M_B$ performed on $A$ and $B$),
 \Eq[A_vs_alpha]{
  H(A|\beta) \leq H(\alpha|\beta),
 }
 where equality holds iff all $\rho(A|\beta=b)$ commute and $\M_A$ is a von
Neumann measurement in the basis where all \hbox{$\rho(A|\beta=b)$} are
diagonal.

\bigskip

Since conditional entropy is meant to express the uncertainty of the state of
subsystem $A$ under the constraints imposed on it by the ``best'' measurement
performed on subsystem $B$, we propose the following

\medskip\noindent{\sc Definition 1.}\quad The conditional entropy of subsystem
$A$, conditioned by subsystem $B$, is
 \Eq[defin]{
 H(A|B) = \inf_{\M_B} H(A|\beta).
 }

\medskip

The following theorem states that, just as in the classical case,
conditioning can only decrease the entropy of a system:

\medskip\noindent{\sc Theorem
1.}
 \Eq[theor1]{
 H(A|B)\leq H(A).
 }

\medskip\noindent{\bf Proof.}\quad
 \Eq{ H(\alpha|\beta)\leq H(\alpha),}
 \Eq{\inf_{\M_B}H(\alpha|\beta)\leq H(\alpha),}
 and
 \Eqaligned{
 H(A|B)&=\inf_{\M_B}\sum_b \Pr\{\beta=b\}\inf_{\M_A}H(\alpha|\beta=b)\\
       &\leq\inf_{\M_A}\inf_{\M_B} H(\alpha|\beta)\leq\inf_{\M_A}H(\alpha)\\
       &= H(A).\halmos
 }

It has been pointed out\cite{yang} that inequality \eq{theor1} follows from the nonnegativity of the entropy defect\cite{levitin3,holevo}. Indeed,
 \Eqaligned{
  &H(A)-H(A|B) =\\
  &\quad = \inf_{\M_B}[-\Tr \rho(A)\log \rho(A)
    +\sum_b\Pr\{\beta=b\}\cdot\Tr\{\rho(A|\beta=b)\log\rho(A|\beta=b)\}],
 }
 where the expression in brackets is a special case of the entropy defect.

\medskip

Note that in classical information theory
$H(\alpha,\beta)=H(\beta)+H(\alpha|\beta)$. However, this equality turns into
an inequality for quantum systems:

\medskip\noindent{\sc Theorem 2}.\quad
 \Eq[theor2]{H(A,B)\leq H(B)+H(A|B).}

\smallskip\noindent{\bf Proof.}\quad By definition \eq{defin}, it suffices to
prove that, for any choice of $\M_B$,
 \Eq[mixed]{
	H(A,B)\leq H(B)+H(A|\beta).
}
 It follows from Naimark's theorem that it is sufficient to prove \eq{mixed}
for the case when $\M_B$ corresponds to a complete set of orthogonal
projectors (a von Neumann measurement).

Let $\{u_{ib}\}$ be the set of orthonormal eigenvectors of the conditional
density matrix $\rho(A|\beta=b)$, and $\{v_b\}$ the set of orthonormal
vectors corresponding to projectors $M_b(B)$. Consider a basis in
$\H=\H_A\otimes\H_B$ formed by vectors $u_{i,b}\otimes v_b$. The joint density
matrix $\rho(A,B)=||\rho_{ib,i'b'}||$ in this basis has the property that
$\rho_{ib,i'b}=\lambda_{ib}\delta_{ii'}$.

Let us introduce a density matrix $\rho'(A,B)$ obtained from $\rho(A,B)$ by
deleting the off-diagonal elements in the basis described above, namely,
 \Eq[rho']{
 \rho'(A,B)=||\rho'_{ib,i'b'}||=||\lambda_{ib}\delta_{ii'}\delta_{bb'}||.
 }
 For each $b$, the conditional density matrix is
 \Eq{
	\rho(A|\beta=b)=
	\left|\left|\frac{\lambda_{ib}\delta_{ii'}}{\sum_i\lambda_{ib}}\right|\right|.
 }
 Also, denote
 \Eq{
 \rho'(B)=\Tr_A\rho'(A,B)=||\delta_{bb'}\sum_i\lambda_{ib}||.
 }
 It is readily seen that, since matrices $\rho'(A,B)$ and $\rho'(B)$ are
diagonal,
 \Eqaligned[Abeta]{
 H(A|\beta) &=-\Tr\rho'(A,B)\ln\rho'(A,B)+\Tr\rho'(B)\ln\rho'(B)\\
            &= -\Tr\rho(A,B)\ln\rho'(A,B)+\Tr\rho(B)\ln\rho'(B).
 }

Consider now the quantum relative entropy (cf.\ \cite{schumacher,vedral})
between states $\rho(A,B)$ and $\rho'(A,B)$,
 \Eq[jointAB]{
 H(\rho(A,B)||\rho'(A,B)) = \Tr\rho(A,B)\ln\rho(A,B)-\Tr\rho(A,B)\ln\rho'(A,B),
 }
 and similarly that between $\rho(B)$ and $\rho'(B)$,
 \Eq[onlyB]{
 H(\rho(B)||\rho'(B)) = \Tr\rho(B)\ln\rho(B)-\Tr\rho(B)\ln\rho'(B).
 }
 It is well known\cite[p.\,13,\,F2]{vedral} that partial tracing reduces
relative entropy; therefore
 \Eq{
 H(\rho(A,B)||\rho'(A,B))\geq HTr_A\rho(A,B)||\\Tr_A\rho'(A,B))=
	H(\rho(B)||\rho'(B)),
 }
 which, by \eq{Abeta}, \eq{jointAB}, and \eq{onlyB}, yields inequality
\eq{mixed}.\halmos

\medskip

According to classical information theory, the information between the
outcomes $\alpha$ and $\beta$ of two POVMs $\M_A$ and $\M_B$ performed on
subsystems $A$ and $B$ is
 \Eq[mutual]{
 I(\alpha;\beta) = H(\alpha)-H(\alpha|\beta).
 }
 In the spirit of Shannon's information theory, we define the mutual
information between two quantum systems as follows:

\medskip\noindent{\sc Definition 2.} The information in subsystem $B$ about
subsystem $A$ (and vice versa) is
 \Eq[about]{I(A;B)=\sup_{\M_A,\M_B} I(\alpha;\beta).}

\medskip

The classical equality \eq{mutual} then turns into an inequality:

\medskip\noindent{\sc Theorem 3.}\quad 
 \Eq[theor3]{I(A;B)\leq H(A)-H(A|B).}

\medskip\noindent{\bf Proof.}\quad From the entropy defect
bound\cite{levitin3,holevo} it follows that for any $\M_B$
 \Eq[proof3]{
 I(A;\beta)= \sup_{\M_A} I(\alpha;\beta)\leq H(A)-H(A|\beta),
 }
 where equality holds iff all $\rho(A|\beta=b)$ commute and $\M_A$ is a von
Neumann measurement in the basis where all $\rho(A|\beta=b)$ are diagonal.

From \eq{proof3},
 \Eqaligned{
 I(A;B) &=\sup_{\M_A,\M_B} I(\alpha,\beta)\\
        &\leq\sup_{\M_B}[H(A)-H(A|\beta)]=H(a)-\inf_{\M_B} H(A|\beta)\\
        &=H(A)-H(A|B).\halmos
 }
 The proposed measures of information and conditional entropy turn out to be
useful in the analysis of correlated (in particular, entangled) quantum
systems, in place of their classical counterparts.

\subsection*{Acknowledgment}

 We would like to thank Dr.\ D.\ Yang (University of
Science and Technology of China) for his thoughtful remarks\cite{yang}.


\end{document}